\documentclass{aa}
\usepackage{graphicx}
\usepackage{longtable}
\usepackage{lscape}
\usepackage{natbib}
\usepackage{amssymb}
\usepackage{amstext}
\usepackage{latexsym} 
\usepackage[sumlimits]{amsmath}
\newcommand{\Ha}{H$\alpha$}			
\newcommand{\HI}{H{\sc i}}			

\begin{document}
   \title{The galaxy population of Abell 1367: photometric and spectroscopic data \thanks{
Based on observations made with the Isaac Newton Telescope and the William Herschel Telescope, operated 
on the island of La Palma by the Isaac Newton Group in the Spanish 
Observatorio del Roque de los Muchachos of the Instituto de Astrof\'\i sica 
de Canarias and the United Kingdom Infra-Red Telescope operated 
on Mauna Kea, Hawaii by the Joint Astronomy Centre. }
   }
   \subtitle{}
   \author{W.~Kriwattanawong \inst{1,2}
   \and C.~Moss \inst{1}\thanks{Deceased 12th May 2010}
   \and P.~A.~James \inst{1} 
   \and D.~Carter \inst{1} 
    } 
          \offprints{W. Kriwattanawong} 
          \institute{
   	  Astrophysics Research Institute, Liverpool John Moores University, Twelve Quays
	  House, Egerton Wharf, Birkenhead CH41 1LD, UK.
          \and
	  Department of Physics and Materials Science, Faculty of Science, Chiang Mai University, Chiang Mai 50200, Thailand.
          }
          \date{Received ; accepted }

\abstract
{}
{ Photometric and spectroscopic observations of the galaxy population of the galaxy cluster Abell 1367 have been obtained, over a field of 34$^{\prime}$$\times$90$^{\prime}$, covering the cluster centre out to a radius of $\sim$2.2 Mpc. Optical broad- and narrow-band imaging was used to determine galaxy luminosities, diameters and morphologies, and to study current star formation activity of a sample of cluster galaxies. Near-infrared imaging was obtained to estimate integrated stellar masses, and to aid the determination of mean stellar ages and metallicities for the future investigation of the star formation history of those galaxies. Optical spectroscopic observations were also taken, to confirm cluster membership of galaxies in the sample through their recession velocities. 
 }
{ $U$, $B$ and $R$ broad-band and H$\alpha$ narrow-band imaging observations were carried out using the Wide Field Camera (WFC) on the 2.5 m Isaac Newton Telescope on La Palma, covering the field described above. $J$ and $K$ near-infrared imaging was obtained using the Wide Field Camera (WFCAM) on the 3.8 m UK Infrared Telescope on Mauna Kea, covering a somewhat smaller field of  0.75 square degrees on the cluster centre. The spectroscopic observations were carried out using a multifibre spectrograph (WYFFOS) on the 4.2 m William Herschel Telecope on La Palma, over the same field as the optical imaging observations. }
{Our photometric data give optical and near-infrared isophotal magnitudes for 303 galaxies in our survey regions, down to stated diameter and $B$-band magnitude limits, determined within $R_{24}$ isophotal diameters. Our spectroscopic data of 328 objects provide 84 galaxies with detections of emission and/or absorption lines. Combining these with published spectroscopic data gives 126 galaxies within our sample for which recession velocities are known.  Of these, 72 galaxies are confirmed as cluster members of Abell 1367, 11 of which are identified in this study and 61 are reported in the literature. H$\alpha$ equivalent widths and fluxes are presented for all cluster galaxies with detected line emission.}
{Spectroscopic and photometric data are presented for galaxies in the nearby cluster Abell~1367, as the first stage of a study of their stellar population and star formation properties.}


\authorrunning{Kriwattanawong et al.}
\titlerunning{The galaxy population of Abell 1367}
\maketitle
%


\section{Introduction}

The impact of the cluster environment on galaxy properties has been extensively investigated during recent decades.  Observational studies starting with the Virgo cluster observations of \citet{ROBI65} have established that cluster galaxies are depleted in \HI\ gas mass when compared with isolated galaxies \citep{HAYN84}.  This deficiency is a strong function of distance from the X-ray centre of the cluster \citep{GAVA89,BOSE94}, and the mass deficit is accompanied by a smaller scale-size of the \HI\ extents for cluster galaxies, relative to optical diameters, than is found for isolated galaxies \citep{WARM88,CAYA94,BOSE02}.  Some cluster disk galaxies appear `anemic' with low disk surface densities of atomic gas \citep{CAYA94}. Unsurprisingly, this gas deficiency is also reflected in lower star formation activity in cluster galaxies.  This was shown by the pioneering \Ha\ work undertaken by \citet{KENN83B}, who found lower star formation rates and redder colours for Virgo Cluster galaxies than for their field counterparts.  Many studies have extended this work to other clusters, with for example \citet{GAVA06} looking at 545 galaxies within the Virgo, Coma and Abell 1367 clusters.  Evidence for spatial truncation of disk star formation, again as traced by \Ha\ emission, has been found by \citet{KOOP98,KOOP04A,KOOP04B} and \citet{KOOP06} 

Many mechanisms have been proposed to explain the observed impacts of the cluster environment on the gas content and star formation activity of cluster galaxies. The interstellar medium can be removed from galaxies by ram pressure stripping, due to the galaxy motion with a velocity of $\sim$1000~km~s$^{-1}$ as it falls through the hot ($\sim10^7-10^8$ K) and dense ($\sim10^{-4}-10^{-3}$ atom/cm$^3$) intra-cluster medium \citep{GUNN72}. Simulations suggest that at least part of the interstellar gas can be stripped out by ram pressure on timescales of a few billion years, comparable to the cluster crossing time \citep{BOSE06}. This has been proposed as a mechanism for the transformation of spiral galaxies into lenticular galaxies in cluster environments. However, ram pressure stripping as the sole mechanism for the transformation of spiral galaxies into lenticulars has received significant criticism. \citet{DRES04} concludes that the age of most lenticulars, and their emergence in relatively low-density environments, argues rather for interaction, merger and accretion processes.  These latter can also thicken disks, and enhance bulge-to-disk ratios, producing galaxies with the observed properties of lenticulars, overcoming another shortcoming of ram pressure stripping scenarios \citep{BOSEEA06}.

Tidal interactions between galaxy pairs moving at low relative velocities can contribute to the enhancement of star formation rates, depending on the gravitational potential between them \citep{BOSE06}. Both simulations \citep{MIHO92,IONO04} and observations \citep{COND82,KEEL85,KENN87} of galaxy-galaxy interactions indicate that they are particularly efficient at triggering star formation in circumnuclear regions via the compression and inflow of gas, with moderate, if any, increase in star formation in the disk. If the encounter between the interacting galaxies is sufficiently close, they can evolve into a merger by dynamical friction.

The molecular gas of infalling galaxies can induce both nuclear and disk star formation, under the tidal effects of the cluster potential \citep{HENR96,FUJI98}. This effect is more effective than ram pressure stripping at enhancing star formation activity. However, it is also possible to remove outer gas, resulting in H$\scriptstyle\rm I$ deficiency of the cluster spiral galaxies \citep{BYRD90}. Supernova-driven winds from the enhanced star formation could also expel the weakly gravitationally bound H$\scriptstyle\rm I$ from the galactic disks \citep{BYRD90,HENR96}. A process termed ``galaxy harassment'' by \citet{MOOR96,MOOR99} involves multiple high speed encounters among galaxies as they move within the cluster potential, and can affect the evolution of cluster galaxies. They suggested that galaxy harassment is very distinct in its effects from other merger processes. This mechanism causes the transformation of low mass disk galaxies to dwarf spheroidal and dwarf elliptical galaxies, and also leaves debris tails. Their simulations have shown that galaxies on elongated orbits undergo more harassment than galaxies on circular orbits. These processes will affect directly the luminosity-weighted mean stellar ages and abundances of heavy elements, which are some of tracers to be used in subsequent papers of this study to trace the evolution of galaxies in a nearby cluster.

One of the most interesting nearby galaxy clusters in terms of its dynamical state is Abell 1367, a spiral-rich system identified in the cluster catalogue of \citet{ABEL58}. The cluster centre is located at $\alpha$(J2000)~$=~11^h44^m29.5^s$, $\delta$(J2000)~$=$~+19$^{\circ}$50$^{\prime}$21$^{\prime}$$^{\prime}$. The average cluster redshift and velocity dispersion are 0.022 and 879~km~s$^{-1}$, respectively \citep{STRU99}. We adopt a value of 1.31$^{\circ}$ for the Abell radius of the cluster, taken from NED, and note that this is completely consistent with the value of 1.29$^{\circ}$ obtained using the $R_{200}$ approximation to the virial radius, from \citet{FINN05}.

This study aims to investigate the evolution of galaxies in the cluster environment, using Abell~1367 as a test case. The main tracers of this study are H$\alpha$ emission, integrated galaxy stellar masses, luminosity-weighted mean stellar ages and metallicities of the cluster members. The results will provide understanding of the effect of the cluster environment on stellar population properties of the galaxies, i.e. chemical enrichment histories and star formation activity.

We present the results of our study of Abell~1367 in two papers. This paper contains a description of the photometric and spectroscopic observations and data analysis, as follows. Photometric observations and data  reduction of optical and near-infrared imaging are described in Sect. 2. Section 3 contains the sample selection made from our imaging. Spectroscopic observations, data reduction and results are described in Sects. 4 and 5. Section 6 explains the extraction of the final photometry from the optical and near-infrared data. Classifications of morphological types and disturbance classifications are detailed in Sect. 7. Section 8 contains a description of the determination of H$\alpha$ emission line properties. A summary of our results is presented in Sect. 9.

Paper II (M. Mouhcine et al., submitted) will provide a description of the determination of luminosity-weighted mean stellar ages and metallicities, along with integrated galaxy stellar masses of cluster members, using the data presented in the current paper. Analysis of those parameters associated with H$\alpha$ emission line properties will also be presented.

\begin{figure}
\hspace*{.5cm}
\includegraphics[width=0.45\textwidth,height=0.52\textheight,angle=0]{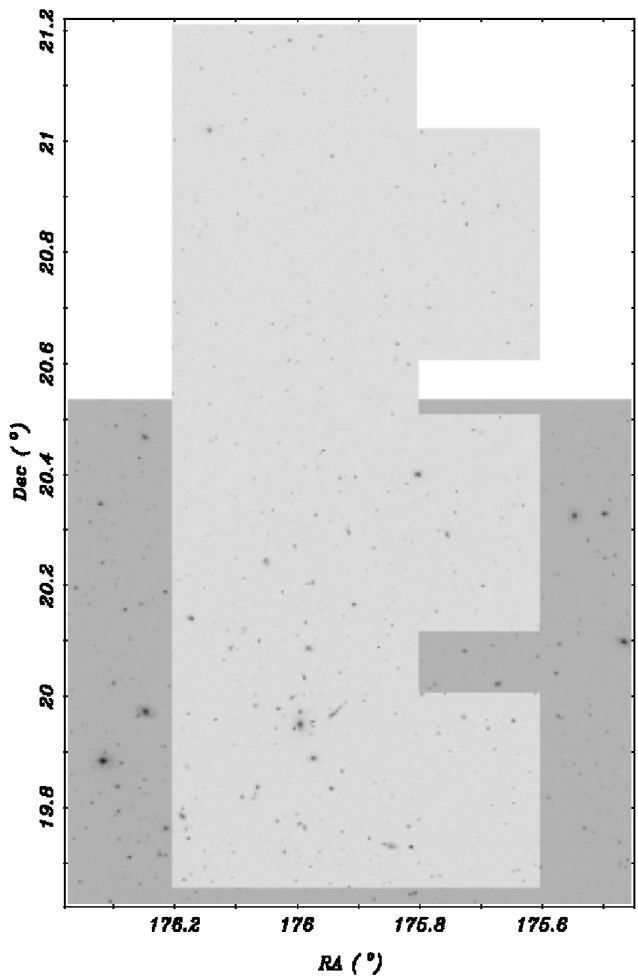}
\caption{The layout of the optical (bright shaded area) and near-infrared (dark shaded area) images covering the central cluster region in J2000 coordinate system. }
\label{UBRJK}
\end{figure}

\section{\label{obs}Photometric Observations and Data Reduction}

$U$, $B$, $R$ and H$\alpha$ imaging was obtained to enable luminosities, colours, diameters and current star formation rates of cluster galaxies to be estimated. Near-infrared $J$- and $K$-band imaging was also obtained, with the aim of estimating integrated stellar masses, and mean stellar ages and metallicities for investigation of star formation history of those galaxies. The data were observed between 2001 and 2006 using several different telescopes as outlined below.

\subsection{UBR and H$\alpha$ data }

$U$- and $B$-band imaging observations of Abell 1367 were carried out on the 2.5 m INT at La Palma. The images covered a 34$^{\prime}$$\times$90$^{\prime}$ field over the area delimited by  $11^h42^m27^s <$ $\alpha$(J2000) $< 11^h44^m53^s$, +19$^{\circ}$38$^{\prime}$50$^{\prime}$$^{\prime}$ $<$ $\delta$(J2000) $<$ +21$^{\circ}$07$^{\prime}$35$^{\prime}$$^{\prime}$. This covers the cluster centre, and extends to a clustercentric radius of about 1.0 Abell  radius ($r_A$) or $\sim$ 2.2 Mpc, assuming $H_0~=$~70~km~s$^{-1}$~Mpc$^{-1}$, to the north direction. These observations were done on 16 April 2001, under the INT service programme. The exposure times in the $U$ and $B$ filters were 3$\times$300 and 3$\times$150 seconds respectively, in each of the 3 pointings needed to cover the target field. These data were obtained using the Wide Field Camera (WFC), which contains 4 EEV 2048$\times$4096 CCDs. Each CCD has 13.5~$\mu$m-squared pixels, corresponding to 0.33$^{\prime}$$^{\prime}$ on the sky.

\begin{table*}
\begin{center}
\caption{\label{intobs} INT observations }
\begin{tabular}{lcccc}
\hline
\hline

Filter Names & Positions (J2000)  & Dates  & No. of Exposures & Exposure Times  \\
 & & &  &  (seconds)  \\
\hline
RGO U   & 11:44:05.6+19:56:20.7  & 16 April 2001 &  3 & 300 \\
RGO U   & 11:44:05.6+20:23:20.7  & 16 April 2001 &  3 & 300 \\
RGO U   & 11:44:05.6+20:50:20.7  & 16 April 2001 &  3 & 300 \\
Kitt Peak B & 11:44:05.6+19:56:20.7  & 16 April 2001 &  3 &  150 \\
Kitt Peak B & 11:44:05.6+20:23:20.7  & 16 April 2001 &  3 &  150 \\
Kitt Peak B & 11:44:05.6+20:50:20.7  & 16 April 2001 &  3 &  150 \\
Harris R   &  11:44:05.6+19:56:20.7 & 20 March 2005 &  1   & 300  \\
Harris R   &  11:44:05.6+20:23:20.7 & 20 March 2005 &  1   & 300  \\
Harris R   &  11:44:05.6+20:50:20.7 & 20 March 2005 &  1   & 300  \\
$[$S$\scriptstyle\rm II$$]$ & 11:44:05.6+19:56:20.7 & 20 March 2005 &  3   & 400 \\
$[$S$\scriptstyle\rm II$$]$ & 11:44:05.6+20:23:20.7 & 20 March 2005 &  3   & 400 \\
$[$S$\scriptstyle\rm II$$]$ & 11:44:05.6+20:50:20.7 & 20 March 2005 &  3   & 400 \\

\hline
\end{tabular}
\end{center}
\end{table*}

Further $R$ and H$\alpha$ images were later observed, again using the INT and WFC. These observations were done on 20 March 2005, and covered the same 3-pointing field as was used for the $U$- and $B$-band imaging. The $R$-band observations used an exposure time of 300 seconds per pointing, whereas the H$\alpha$ images used exposure times of 3$\times$400 seconds per pointing. The H$\alpha$ imaging was taken through what is nominally a [S{\sc ii}] filter, centred on 6725~\AA, which was used because it covers the redshifted H$\alpha$ emission line from galaxy members of the cluster Abell 1367, with a central recession velocity of 6595~km~s$^{-1}$. The transmission of this filter falls to 50\% of the peak value for wavelengths corresponding to H$\alpha$ redshifted by 6040 and 9430~km~s$^{-1}$ at the short and long wavelength ends respectively, and to 10\% for velocities of 5200 \& 10500~km~s$^{-1}$.  The high end of these ranges accommodates all likely cluster members, but the low velocity cutoff only extends to $\sim$0.6 times the velocity dispersion ($\sigma$) with good sensitivity, and galaxies beyond 1.6~$\sigma$ are very unlikely to be detected in our narrow-band imaging. Thus some true cluster members may be missed from the H$\alpha$ emission line catalogue, if they lie in the low-velocity wing of the cluster distribution. Table $\ref{intobs}$ gives a summary of all the INT observations.

The layout of the optical image field is shown as the lighter shaded field elongated N - S in Fig. $\ref{UBRJK}$. Each pointing of INT/WFC covers a field of view with an overall envelope of 34$^{\prime}$$\times$34$^{\prime}$, but with a missing corner and 1$^{\prime}$ gaps between the 4 CCD chips. The observation was done in 3 pointings, with the first pointing on the cluster central area and the other two fields offset to the North. There are 2 overlap areas; galaxies in these regions were used to propagate the photometric zero-points between the 3 pointings.

\subsection{\label{jk}JK data}

\begin{table*}
\begin{center} 
\caption{\label{ukirtobs} UKIRT observations at a tile position of 11h43m39.6s+20$^{\circ}$05$^{\prime}$40$^{\prime}$$^{\prime}$ (J2000) }
\begin{tabular}{c|lccc}
\hline
\hline

Filter Names & Positions & Dates  & No. of Exposures & Exposure Times  \\
 & & &  &  (seconds)  \\
\hline

J & tile NE & 13 May 2006 & 2 & 20$\times$10 \\
J & tile SE & 13 May 2006 & 2 & 20$\times$10 \\
J & tile NW & 13 May 2006 & 2 & 20$\times$10 \\
J & tile SW & 13 May 2006 & 1 & 20$\times$10 \\
K & tile NE & 13 May 2006 & 2 & 20$\times$10 \\
K & tile SE & 13 May 2006 & 2 & 20$\times$10 \\
K & tile NW & 13 May 2006 & 2 & 20$\times$10 \\
K & tile SW & 13 May 2006 & 2 & 20$\times$10 \\
K & tile NE & 10 June 2006 & 1 & 20$\times$10 \\
K & tile SE & 10 June 2006 & 1 & 20$\times$10 \\
K & tile NW & 10 June 2006 & 1 & 20$\times$10 \\
K & tile SW & 10 June 2006 & 1 & 20$\times$10 \\

\hline
\end{tabular}

\end{center}
\end{table*}

Near-infrared images were obtained from the 3.8~m United Kingdom Infrared Telescope (UKIRT) on Mauna Kea, Hawaii with the wide field camera (WFCAM) using 4 Rockwell Hawaii-II (HgCdTe 2048$\times$2048) arrays. These $J$- and $K$-band images covered 0.75 square degrees with 0.4$^{\prime}$$^{\prime}$pixels, focussed on the cluster centre field that contains the majority of our sample galaxies. There are usually 4 sub-pointings per tile which covers a 53$^{\prime}$$\times$53$^{\prime}$ field of view. For this study, the $J$- and $K$-band images covered the region $11^h41^m46.3^s <$ $\alpha$(J2000) $< 11^h45^m34.0^s$, +19$^{\circ}$38$^{\prime}$05$^{\prime}$$^{\prime}$ $<$ $\delta$(J2000) $<$ +20$^{\circ}$31$^{\prime}$27$^{\prime}$$^{\prime}$ as shown in Fig. $\ref{UBRJK}$.  The $J$ observations were made on 13 May 2006, and there is one tile with only 3 sub-pointings and hence somewhat lower sensitivity than the rest of the surveyed area. The $K$-band images were obtained for 2 tiles in the same night as the $J$-band images, with one more tile being observed on 10 June 2006, as shown in table $\ref{ukirtobs}$. All near-infrared images were taken with exposure times of 20$\times$10 seconds.

\subsection{Data reduction}

Both optical and near-infrared images were initially reduced by the Cambridge Astronomical Survey Unit (CASU) of the Institute of Astronomy, Cambridge University.  $U$, $B$, $R$ and H$\alpha$ images were processed using the pipeline constructed by CASU for the INT Wide Field Camera, and $J$ and $K$ images were processed using the equivalent pipeline for UKIRT/WFCAM images. Although a preliminary astrometric solution was calculated by the pipeline, the accuracy for our INT images was only about 2$^{\prime\prime}$, so an additional astrometric calibration was applied using  the Starlink ASTROM package, applied to the $B$-band images. This astrometry was entirely satisfactory for determining galaxy positions, with the fit to the astrometric star positions having residuals of typically about 0.2$^{\prime\prime}$.

\subsubsection{\label{zero}Photometric zero-point calibrations }

The imaging observations presented here were obtained on nights with average to good seeing (1.3 - 1.0$^{\prime\prime}$) and probably photometric conditions.  However, there was some cloud at the start of the night in which the $U$- and $B$-band imaging was taken. Given this, and given the availability of high-quality published photometry for many of the brighter galaxies in our fields, the latter were used for photometric calibration and the determination of zero-points in each passband.  Systematic variations in the sky level (due to, e.g., flat-fielding errors) were estimated using 5 sky regions around each of the calibrator galaxies. Photometric errors including photon shot noise from the objects and from the sky background, extracted from the GAIA package, were averaged and used as one source of the errors of calibrator zero-points. The standard errors of the instrumental magnitudes obtained using 5 sky regions were taken as errors resulting from sky level uncertainty. These 2 sources of error were combined to obtain overall random magnitude errors of each calibrator galaxy. The instrumental magnitudes were averaged from the 5 measurements and subtracted from the published magnitudes to obtain calibrator zero-points. The pointing zero-points are the mean of the calibrator zero-points for that pointing. The standard errors of calibrator zero-points and the average of the random errors of the calibrator magnitudes were combined to give the total errors on the zero-points. These total errors will be used as the systematic errors of magnitudes for each band.

Magnitudes quoted in specific apertures were preferred for this calibration. This requires sources of published galaxy photometry using defined apertures in each band. For example, the zero-points of the $U$- and $B$-band images used multiple fixed aperture sizes, clearly specified in arcseconds, by \citet{BUTA96}. No such aperture photometry was found for $R$-band calibration, and here image zero-points were calculated using photometric data from \citet{DEVA88} and \citet{TAYL05}, which are estimated total magnitudes in the Johnson $R$ filter. As a result, the apertures for the $R$-band images were made as large as possible to obtain total fluxes of those galaxies without contamination from other objects. $J$- and $K$-band magnitudes of calibrating galaxies were published by the Two Micron All Sky Survey \citep[2MASS;][]{JARR00,JARR03}. Three types of magnitudes from 2MASS were considered for our calibration purposes: total magnitudes, magnitudes within the $\mu_K=20.0$ isophote and those measured in 14 arcsecond diameter apertures. Because 14 arcsecond diameter is a well-defined aperture, this was used to measure fluxes of the galaxies to calibrate zero-points for the near-infrared bands. Flux zero-point calibration will be described in section 8.2.

\subsubsection{\label{GK}Galactic extinction and K-correction}

The measured magnitudes of galaxies in all bands are always fainter than their intrinsic values due to the effect of Galactic extinction. Although this effect is small for this cluster, which is at a Galactic latitude of 73$^{\circ}$, E($B-V$)$=$0.023, it still affects galaxy colours to a non-negligible extent. In particular, the ($B-K$) and ($J-K$) colours are important parameters for the stellar population analysis to be presented in later papers, and so these magnitudes and colours need to be corrected to achieve the accuracy needed to fit to models. The Galactic extinction corrections for all bands were obtained from NED, derived by \citet{SCHL98}, using the $R_V~=~3.1$ extinction laws of \citet{CARD89} and \citet{ODON94}.

Furthermore, measured galaxy colours are also changed from the intrinsic colours, due to redshift effects. Abell 1367 is a nearby cluster, which means the effects are small, but it is still the case that detected fluxes correspond to slightly shorter emitted wavebands, corresponding to the redshift $z=$0.022. This effect can cause colours to become bluer or redder, depending on which colours are considered. K-corrections to remove this effect for optical and near-infrared wavebands were calculated by \citet{POGG97}, who also considered the variations in the required corrections depending on galaxy morphological type (and hence intrinsic colour). For this study, 2 different corrections from those produced by \citet{POGG97} were applied, one corresponding to early types (applied to elliptical and lenticular galaxies) and one to late types (applied to spiral and irregular galaxies). The corrections applied are 0.077, 0.107, 0.023, -0.007 and -0.033~mag for early-type galaxy in the {\it UBRJK} bands, respectively; and 0.065, 0.072, 0.012, -0.014 and -0.035~mag for late-type galaxies. 

However, Galactic extinction and K-corrections are not applied to the data in the table $\ref{Photom}$. These corrections are required in the analysis of stellar properties of the cluster galaxies, using optical/near-infrared colours, and will thus be applied in analysis presented in later papers.

\subsubsection{\label{isoph}Isophotal apertures }

Galaxy diameters are normally determined at isophotal surface brightnesses. In this study, the best-fit ellipse corresponding to a surface brightness of 24 magnitude$\slash$arcsecond$^2$ in the $R$ band, $R_{24}$, was determined using a standard surface photometry package and adopted as our standard galaxy aperture. Thus the isophotal semi-major axis, ellipticity and position angle values for each galaxy were used to set the aperture sizes for measurements of photometric magnitudes and fluxes in all bands. Sky background uncertainty will considerably affect the surface photometry, especially for faint magnitude levels. Thus close attention was paid to sky subtraction and associated uncertainties.

\section{Sample Selection  }
\label{SamSel}

\begin{figure}
\centering
\includegraphics[width=0.33\textwidth,height=0.35\textheight,angle=-90]{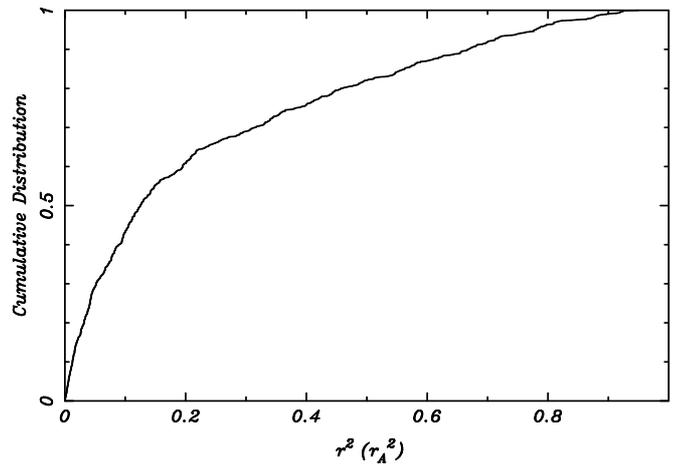}
\caption{The cumulative distribution of galaxy number versus the square of distance from the cluster centre.  }
\label{cumr}
\end{figure}

Two criteria were used for the initial selection of galaxies in our survey region: a $B$-band magnitude $m_B$ brighter than 21.0, and an object size greater than 7.0 pixels ($\sim$~2.3 arcsecond) full width half maximum (FWHM), determined by using the SExtractor package \citep{BERT96}, adopting default parameters. These two criteria were applied to the full list of objects found by SExtractor over the whole field, with the first of them setting the faint magnitude limit of our final sample. As a first step in exploring the likely contamination from background objects, the surface number densities of objects in this magnitude-limited parent sample were used to estimate the effective cluster radius and the probable fraction of cluster members inside this radius. Figure~\ref{cumr} shows the cumulative radial number counts of objects satisfying the above apparent magnitude and diameter criteria, plotted against the square of the distance from the cluster centre, such that a constant surface density background will give a linear rise in counts.  The plot shows the rapid central rise in counts due to true cluster members, which changes in a well-defined manner to an approximately linear rise beyond $r^2\sim0.2$ on the plot, corresponding to a radius of 0.4--0.5 times the Abell radius. This implies that the extended objects distribute uniformly outside a radius of about 0.5 $r_A$, indicating that they are dominated by background galaxies, and that 0.5 $r_A$ is an appropriate limit for the region dominated by true cluster galaxies. 

The final selected sample of cluster galaxy candidates, by apparent magnitude and diameter, comprises 303 galaxies. The true nature of these galaxies, whether cluster or background, will be further investigated in this paper using literature and newly obtained spectroscopic redshifts (see section $\ref{SpecR}$).

\begin{table*}
\begin{center}
\caption{\label{wyfobs} WHT/WYFFOS observations}

\begin{tabular}{lcccc} 
\hline
\hline

Positions (J2000) & Dates & No. of Exposures & Exposure Times & No. of Target Objects\\
 & &  & (seconds) \\
\hline

11 43 37.53 +20 09 31.4  & 30 April 2002 &  4 & 1800 & 36 \\
11 43 39.73 +19 56 05.3  & 25 March 2003 &  4 & 1800 & 72 \\
11 43 40.07 +20 23 06.3  & 26 March 2003 &  4 & 1800 & 76 \\
11 43 39.79 +20 50 25.9  & 26 March 2003 &  4 & 1800 & 71 \\
11 43 40.33 +19 55 44.2  & 26 March 2003 &  3 & 600 & 73 \\

\hline
\end{tabular}

\end{center}
\end{table*}

\section{Spectroscopic Observations and Data Reduction}

 The spectroscopic data were obtained using the WYFFOS multifibre spectrograph on the 4.2 m WHT. The total area covered approximately the same field as the INT$\slash$WFC imaging observations. The WHT/WYFFOS observations were done in 5 pointings on 3 nights, 30~April~2002, 25 March 2003 and 26 March 2003 as shown in table $\ref{wyfobs}$. The exposure times were 4$\times$1800 seconds for the first 4 pointings and 3$\times$600 seconds for the last pointing. All pointings used neon arc lamps for wavelength calibration.

For these observations, the R600R grating was used, giving a dispersion of approximately 3.0~$\rm\AA$~pixel$^{-1}$, for a total spectral coverage of 3080 $\rm\AA$ with a TEK 1024$\times$1024 CCD. The spectra were centered on 6000 $\rm\AA$, covering H$\beta$~$\lambda$~4861, [O$\scriptstyle\rm III$]~$\lambda$~4959, [O$\scriptstyle\rm III$]~$\lambda$~5007, H$\alpha$~$\lambda$~6563, [N$\scriptstyle\rm II$]~$\lambda$~6548~+~6583, [S$\scriptstyle\rm II$]~$\lambda$~6717 and  [S$\scriptstyle\rm II$]~$\lambda$~6731 emission lines, plus Mg$\scriptstyle\rm I$~$\lambda$~5175 and NaD~$\lambda$~5893 absorption lines.

The spectroscopic sample was selected to discriminate between cluster galaxies and background galaxies or AGN, using a criterion of $B~<~22$, over the same sky area as the photometric data. Three hundred and twenty eight target objects were observed as shown in table $\ref{wyfobs}$. 

The spectroscopic data reduction was mainly done using IRAF packages. First, bias frames were median-combined and subtracted from the other frames such as science, neon arc and offset sky frames. Science frames were median-combined for each pointing which efficiently removed cosmic rays. The science, neon arc and offset sky images were used as input files for the WYFRED package, which is a special package for the extraction of multifibre spectra produced by WHT/WYFFOS. WYFRED was run including automatic aperture number identifications, to match up spectral tracks on the CCD with fibres and hence target objects. During the running of this program, the aperture numbers were identified and a neon arc frame was used for wavelength calibration in each fibre. Finally, sky subtraction was performed, one-dimensional spectra were extracted and redshifts were determined by using the ONEDSPEC package.

\begin{figure}
\centering
\includegraphics[width=0.35\textwidth,height=0.35\textheight,angle=-90]{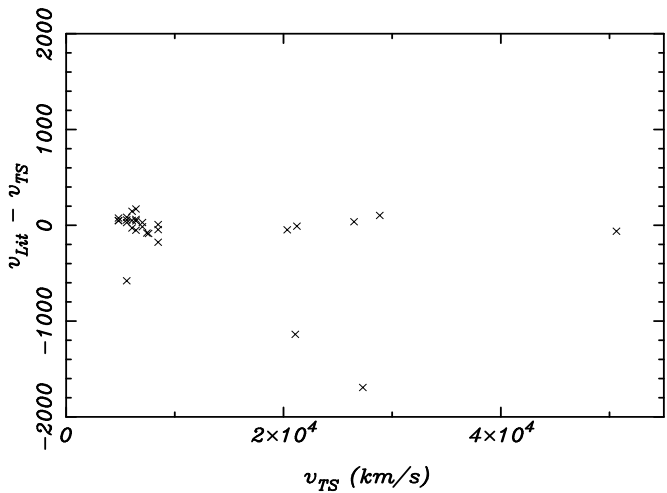}
\caption{Comparison of radial velocities between the literature ($v_{Lit}$) and this study ($v_{TS}$). Many galaxies have published redshifts from several papers, listed in table $\ref{complst}$.}
\label{v.Lit}
\end{figure}

\begin{table*}
\begin{center}
\caption{\label{complst}Comparison of radial velocities between the literature ($v_{Lit}$) and this study ($v_{TS}$).   }
\begin{tabular}[]{lccccccccc} 
\hline
\hline

SDSS & $v_{TS}$ & &&& $v_{Lit}$ &&&& \\
& (km \rm s$^{-1}$) & &&& (km \rm s$^{-1}$) &&&& \\

\cline{3-10}
&& &&& Ref.&&&& \\
&& (1) & (2) & (3) & (4) & (5) & (6) & (7) & (8) \\
\hline

 J114238.29+194718.1 &27303& - &25610& - & - & - & - & - & - \\
 J114239.31+195808.1 &7426& - & - & - & - & - &7345& - & - \\
 J114240.36+195628.0 &21084& - &19946& - & - & - & - & - & - \\
 J114240.34+195717.0 &7586& - & - &7501& - & - & - & - & - \\
 J114245.43+195042.5 &21234& - & - & - & - & - &21225& - & - \\
 J114313.09+200017.3 &7015& - & - & - & - & - & - &7044&7000\\
 J114335.10+202928.6 &50637& - & - & - & - & - &50575& - & - \\
 J114354.06+195952.2 &28857& - & - & - & - & - &28960& - & - \\
 J114357.50+195713.6 &6068&6214& - & - &6119& - & - & - &6040\\
 J114358.23+201107.9 &6425&6471& - & - &6487& - &6595& - &6373\\
 J114442.95+200440.4 &26495& - & - & - & - & - &26532& - & - \\
 J114447.79+194624.3 &8470& - & - & - &8478&8425& - & - &8293\\
 J114447.95+194118.6 &4819&4895& - & - &4866& - & - & - &4869\\
 J114448.90+194828.8 &20343& - & - & - & - & - &20296& - & - \\
 J114449.16+194742.2 &5579&5607& - & - &5663&5635& - & - &5000\\

\hline
\end{tabular}
\end{center}

\begin{tabular}{l}

\\

\it Notes to Table $\ref{complst}$: 
\rm Column 1: SDSS identifications. Column 2: recession velocities from this study. 
Columns 3-10: recession \\velocities from the literature as follows:  Smith et al. (2004), Cortese et al. (2004), Cortese et al. (2003), Rines et al. (2003), \\Gavazzi et al. (2003b), Sullivan et al. (2000), Haynes et al. (1997) and Zabludoff et al. (1990), respectively.

\end{tabular}

\end{table*}

\section{\label{SpecR}Spectroscopic Results}

Most cluster members had recession velocities determined from the H$\alpha$~$\lambda$~6563,  [N$\scriptstyle\rm II$]~$\lambda$~6548 and  [N$\scriptstyle\rm II$]~$\lambda$~6583 lines. For some galaxies we also found  [S$\scriptstyle\rm II$]~$\lambda$~6717 and  [S$\scriptstyle\rm II$]~$\lambda$~6731 emission, and NaD~$\lambda$~5893 absorption lines. Many background galaxies out to $z$~$\sim$~0.57 were strongly confirmed by not only the above lines but also H$\beta$~$\lambda$~4861, [O$\scriptstyle\rm III$]~$\lambda$~4959 and [O$\scriptstyle\rm III$]~$\lambda$~5007 emission, and  FeI~$\lambda$~5269 and MgI~$\lambda$~5175 absorption lines. Moreover, shorter wavelength spectral line features were also found such as [O$\scriptstyle\rm II$]~$\lambda$~3727 emission, and G-band~$\lambda$~4304 and CaH~$\lambda$~3968 absorption lines for distant galaxies. The redshifts from several lines were averaged for each galaxy. 

There are 84 out of 328 objects for which spectral lines were clearly detected, with many of the spectroscopic targets being too faint to result in spectral line detections using WHT/WYFFOS. Redshifts were determined for all galaxies with detected lines, as shown in table $\ref{SpecCat}$. Of these, 80 galaxies have detected emission lines, and some of them also show absorption lines, whereas 4 galaxies were found with only absorption lines. Furthermore, many galaxies' redshifts were collected from published papers as noted at the bottom of table $\ref{complst}$. Fifteen galaxies for which we have spectroscopic data also have redshifts in the literature. Some galaxies have redshifts from several different papers, as shown in table $\ref{complst}$ and Fig. $\ref{v.Lit}$. Absorption line recession velocities were calculated from estimates of individual line centre positions, and not using the cross-correlation technique.  However, an analysis of the 2 objects with both emission and absorption lines showed good overall agreement in velocities. The velocities of J114324.70+194831.8 are of 13193~km~s$^{-1}$ and 13320~km~s$^{-1}$ for H$\alpha$ and [N$\scriptstyle\rm II$]~$\lambda$~6583 emission lines, and 13241~km~s$^{-1}$ for NaD absorption line. The velocities of J114449.16+194742.2 are 5626~km~s$^{-1}$ and 5591~km~s$^{-1}$ for H$\alpha$ and [N$\scriptstyle\rm II$]~$\lambda$~6583 emission lines and 5493~km~s$^{-1}$ and 5605~km~s$^{-1}$ for MgI and NaD absorption lines.

Our values show generally good agreement with published redshifts. There are 3 outlying galaxies. The first is SDSS J114449.16+194742.2, which has a radial velocity of 5579~km~s$^{-1}$, as measured in this study. There are 4 published papers with measured recession velocities for this galaxy. Three of them are in close agreement, with velocities of 5607, 5663 and 5635~km~s$^{-1}$ from \citet{SMIT04}, \citet{RINE03} and \citet{GAVA03B}, respectively, whereas the other reported value was 5000~km~s$^{-1}$, from \citet{ZABL90}. Thus all these studies apart from the last agreed with our value of $\sim$ 5600~km~s$^{-1}$. The other two discrepant velocities are for SDSS J114238.29+194718.1 and SDSS J114240.36+195628.0, both of which have recession velocities published in the same paper \citep{CORT04}. The published velocities differ from our values by 6 and 5$\%$, respectively. However, for the other 13 galaxies, our results are in good agreement with literature values taken from several papers. Many galaxies have published redshifts from more than one paper, as listed in Table $\ref{complst}$.

\begin{figure}
\centering
\includegraphics[width=0.35\textwidth,height=0.35\textheight,angle=-90]{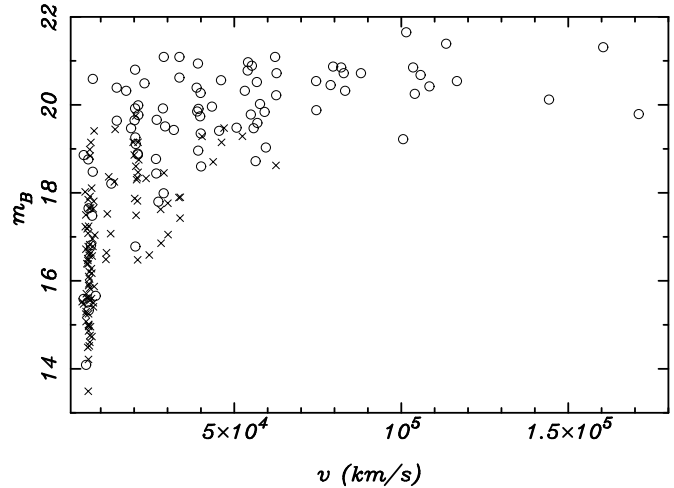}
\caption{The $B$-band magnitude distribution of galaxies with known redshifts: this work (open circle) and previous papers (cross).}
\label{vB}
\end{figure}

After our spectroscopic results were combined with published recession velocities, it was found that the cluster members' radial velocities are between 4455 and 8483~km~s$^{-1}$, scattered over a range 2.5 times the velocity dispersion (879~km~s$^{-1}$) about the average cluster velocity, 6595~km~s$^{-1}$ \citep{STRU99} as shown in Fig. $\ref{vB}$. The $B$-band apparent magnitude distribution of spectroscopically-confirmed cluster members scatters mostly between 14 and 20. All galaxies within our survey field that are brighter than $m_B=$16 magnitudes are cluster members, whilst the distant galaxies distributed between $m_B=$16 and 22. Almost all galaxies fainter than $m_B=$ 20 are background galaxies. The background galaxies identified in this study have radial velocities scattered between 12,000 and 160,000~km~s$^{-1}$. There is strong evidence for a background cluster with an average velocity of $\sim$ 20,000~km~s$^{-1}$ within this field. Several background galaxies have a redshift of $\sim$ 0.34 (just over 10$^5$~km~s$^{-1}$ on Fig. \ref{vB}), which can result in their [O$\scriptstyle\rm III$]~$\lambda$~5007 line emission being confused with H$\alpha$ from Abell 1367 cluster members. This is an important consideration when using emission within narrow-band filters as a selection criterion for cluster membership.

\begin{figure}
\centering
\includegraphics[width=0.35\textwidth,height=0.35\textheight,angle=-90]{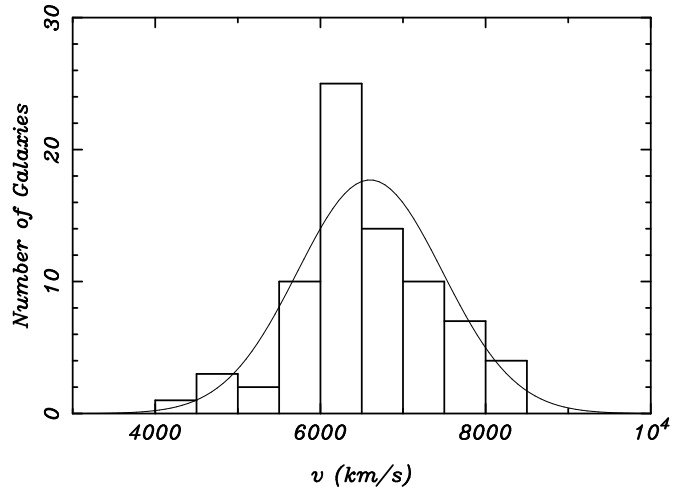}
\caption{The histogram of the cluster members, identified by spectroscopic redshifts, is overlaid by a Gaussian normal distribution, using the mean redshift and velocity dispersion, reported by \citet{STRU99}. }
\label{barv}
\end{figure}

Eleven of the 84 galaxies for which we have spectroscopic data are members of Abell 1367, whereas the others are background galaxies. The spectroscopic recession velocities are combined with literature data to verify cluster membership for the 303 selected galaxies from  section $\ref{SamSel}$; we have 126 known redshifts for these galaxies. It was confirmed that 72 galaxies, 11 galaxies from this study and 61 galaxies from literature, are associated with the cluster Abell~1367, whereas 54 galaxies are background objects. A histogram of cluster members, including spectroscopic redshifts from this work and from previous studies, is plotted in Fig. $\ref{barv}$. The overlaid curve shows a normal distribution using the mean redshift and velocity dispersion reported by \citet{STRU99}.

\section{Optical/Infrared Photometry  }
\label{  }

The $R_{24}$ isophotal apertures, defined in section $\ref{isoph}$, are used to determine $UBRJK$ isophotal magnitudes for our galaxy sample on the Vega magnitude system, listed in table $\ref{Photom}$. Column 1 lists SDSS identifications. Major and minor diameters at the $R_{24}$ isophotes are shown in columns 2 -- 3. Columns 4 -- 8 list the $UBRJK$ magnitudes with errors. There are two kinds of the errors; random and systematic errors. The random magnitude errors occur from photon shot noise of the objects and from the uncertainty of the sky background as discussed in section $\ref{zero}$.  The main systematic error is the uncertainty of zero-points, also mentioned in section $\ref{zero}$.

The comparison of UBRJK magnitudes between the literature and this study are separately calculated for R and UBJK magnitudes. The literature $R$-band magnitudes were derived from the SDSS and hence were originally measured in the SDSS $r$ filter system.  Athough there is some scatter between these and our $R$-band measurements, there is no evidence for an overall offset, as is shown by the mean offset of just 0.04~$\pm$~0.03 mag. For the UBJK magnitudes, the comparison was done with observations of a subset of our galaxies that are contained in the GOLDMine database (Gavazzi et al. 2003a).  One complication is that the present study uses $R_{24}$ isophotal diameters, whereas GOLDMine uses $B_{25}$.  For bright galaxies these diameters are in general comparable in size, but we found that for fainter galaxies, which can have much bluer colours, the $R_{24}$ diameters tend to be substantially smaller, often by a factor of 2.  This clearly precludes any direct photometric comparison for these galaxies.  However, for the 9 galaxies where the $R_{24}$ and $B_{25}$ galaxies agree within $\pm$~30~$\%$, an approximate comparison can be done.  For these galaxies, there is marginal evidence that this study finds magnitudes fainter than those quoted by GOLDMine, with the photometry offsets are 0.04~$\pm$~0.03 mag in the U band, 0.14~$\pm$~0.06 mag in the B band, 0.19~$\pm$~0.12 mag in the J band and 0.18~$\pm$~0.12 mag in the K band, in each case. But the offsets are not much larger than the standard errors and, due to the remaining aperture differences, a small systematic offset is not surprising.

\section{The Morphological Types}
\label{  }

In this study, our $B$-band images were used to classify galaxy types in the Hubble system, using de Vaucouleurs's T-type. Disturbance, meaning galaxies with clearly asymmetric or peculiar morphologies, was also identified. There was, however, a problem of edge-on spirals, which could not be clearly classified into sub-types of spiral galaxies. These galaxies were classified as S types and identified as T-type 11. As a result, our galaxies were classified into 13 T-types. These cluster members are quite bright, and generally the morphological classifications were clear and unambiguous. The morphological types, T-types and disturbance classifications are listed in table $\ref{PhotCat}$ (columns 2 -- 4).   

Table $\ref{Typetab}$ shows comparison of the morphological types between this study and the literature, collected from the GOLDMine database \citep{GAVA03A}. There are 31 of 72 cluster galaxies, recorded in the GOLDMine database. If the `Pec' type of the GOLDMine database is considered as equivalent to the `Disturbed' type of this study, we find that 23 galaxies show the same types, and 5 galaxies are similar (i.e., SB0 in this study and S0 in literature, or S0/a in this study and S0 in literature). Given the subjective nature of galaxy classifications, this is a strikingly good degree of agreement.

\begin{table}
\begin{center}
\caption{\label{Typetab} Comparison of the morphological types between this study and the literature. }
\vspace{\baselineskip}
\begin{tabular}{@{}llcl@{}} 
\hline
\hline
\multicolumn{1}{l}{SDSS } &
\multicolumn{2}{c}{This study } &
\multicolumn{1}{l}{Literature} \\
\cline{2-3}
& Type & Disturbance & Type \\

\hline

  J114256.45+195758.3 	&	                 Scd                 	&	                 Disturbed                 	&	Pec	\\
  J114313.09+200017.3	&	S,Pec  	&	                 Disturbed                 	&	Pec	\\
  J114324.55+194459.3 	&	                 Sa                 	&	                         -                         	&	Sa	\\
  J114343.96+201621.7	&	                 S0/a                 	&	                         -                         	&	S0	\\
  J114349.07+195806.4	&	                 Scd                 	&	                 Disturbed                 	&	Pec	\\
  J114356.42+195340.4	&	                 E                 	&	                         -                         	&	E	\\
  J114357.50+195713.6	&	                 S0                 	&	                         -                         	&	S0	\\
  J114358.23+201107.9	&	                 S0                 	&	                 Disturbed                 	&	Sbc	\\
  J114358.95+200437.2	&	                 Sa                 	&	                         -                         	&	Sa	\\
  J114359.57+194644.2 	&	                 SB0                 	&	                         -                         	&	S0	\\
  J114401.94+194703.9	&	                 Sc                 	&	                 Disturbed                 	&	Pec	\\
  J114402.15+195659.3 	&	                 E                 	&	                         -                         	&	E	\\
  J114402.15+195818.8	&	                 E                 	&	                         -                         	&	E	\\
  J114403.21+194803.9	&	                 E                 	&	                         -                         	&	E	\\
  J114405.46+195945.9 	&	                 SB0                 	&	                         -                         	&	S0	\\
  J114407.64+194415.1	&	                 E                 	&	                         -                         	&	E	\\
  J114416.48+201300.7	&	                 E                 	&	                         -                         	&	E	\\
  J114417.20+201323.9 	&	                 Sa                 	&	                 Disturbed                 	&	Sa	\\
  J114420.42+195851.0	&	                 E/S0                 	&	                         -                         	&	S0	\\
  J114420.74+194933.3	&	                 SB0                 	&	                         -                         	&	S0a - S0/Sa	\\
  J114425.10+200628.1	&	                 E                 	&	                         -                         	&	E	\\
  J114425.12+194941.0 	&	                 S0                 	&	                         -                         	&	S0	\\
  J114425.92+200609.6 	&	                 S0                 	&	                 Disturbed                 	&	Pec	\\
  J114428.36+194406.6	&	                 E                 	&	                         -                         	&	E	\\
  J114430.55+200436.0	&	                 S0                 	&	                         -                         	&	S0	\\
  J114446.62+194528.3	&	                 S0                 	&	                         -                         	&	E	\\
  J114447.03+200730.3	&	                 Sbc                 	&	                 Disturbed                 	&	Sab	\\
  J114447.44+195234.9	&	                 S0                 	&	                         -                         	&	S0	\\
  J114447.79+194624.3	&	                 Sc                 	&	               Disturbed                 	&	Pec	\\
  J114447.95+194118.6	&	                 SB0/a                 	&	                         -                         	&	Sa	\\
  J114449.16+194742.2	&	                 Sa                 	&	                         -                         	&	Sa	\\

\hline

\end{tabular}
\end{center}
\end{table}

\section{H$\alpha$ Emission  }
\label{  }

H$\alpha$ emission is a powerful tracer for the study of current star formation activity in galaxies, and can be determined from H$\alpha$ filter images from which an appropriate subtraction of continuum emission has been made \citep{KENN83A,KENN94,GAVA06}. The H$\alpha$ emission line at 6563 $\rm\AA$ lies between neighbouring [N$\scriptstyle\rm II$] lines (6548 and 6583 $\rm\AA$), which are all transmitted through the filter this study uses. Thus fluxes presented in this study correspond to  H$\alpha$ emission plus the associated [N$\scriptstyle\rm II$] lines; appropriate corrections for this \citep{KENN83A,KENN98} need to be applied to these fluxes when deriving, for example, star formation rates.

\subsection{Calculation of H$\alpha$ equivalent widths}

The H$\alpha$ equivalent width, $EW$, is a parameter that can be used to study the star formation activity of galaxies normalised by luminosity. The $EW$ divides the star formation activity H$\alpha$ line fluxes by the continuum fluxes of the older stellar populations in the galaxy. It can be determined by equation $\ref{EW2}$ \citep{GAVA06}:

\begin{equation}
 EW = \frac{\int T_n(\lambda)d\lambda}{T_n(6563(1+z))} \times \frac{C_{H\alpha}}{C_C},
\label{EW2}
\end{equation}

where $C_C$ is continuum flux counts, obtained in this study from $R$-band images, $C_{H\alpha}$ is the continuum-subtracted H$\alpha$ emission line flux counts, and $T_n(\lambda)$ is the transmissivity of the narrow band filter. The narrow band filter, [S$\scriptstyle\rm II$] 6725/80, used in this study has good transmissivity in the wavelength range between 6695~and~6769~$\rm\AA$. 

\begin{figure}
\centering
\vspace*{.6cm}
\includegraphics[width=0.35\textwidth,height=0.35\textheight,angle=-90]{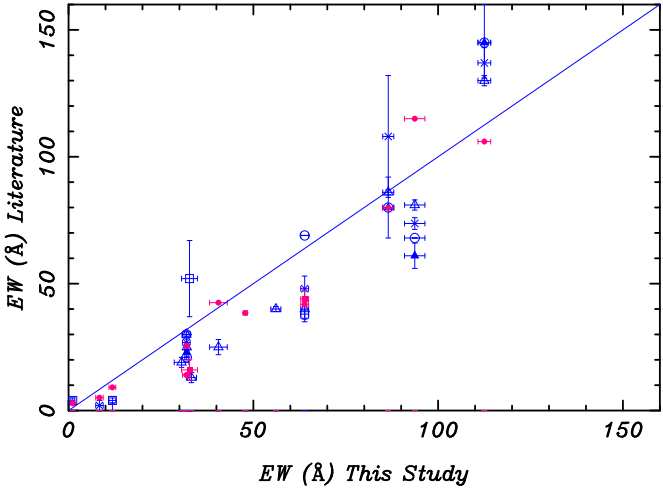}
\caption{Comparison of the H$\alpha$ equivalent widths between the literature and this study. The symbols are listed in table $\ref{EWLit}$. }
\label{EW}
\end{figure}

Figure $\ref{EW}$ shows a correlation between the H$\alpha$ equivalent widths from the literature and this study; the diagonal line shows the one-to-one correlation. The symbols indicate which of the 9 literature sources listed in Table $\ref{EWLit}$ provided the data for this comparison. Most of these papers published H$\alpha$ + [N$\scriptstyle\rm II$] equivalent widths and fluxes, represented by blue symbols. Only the papers numbered 8 and 9 in the list determined H$\alpha$ emission separate from [N$\scriptstyle\rm II$], and these are represented by red symbols.

\begin{table*}
\centering
\caption{\label{EWLit}Papers containing H$\alpha$ equivalent widths and fluxes of galaxy members in Abell 1367. The symbols are refer to those used in Figs. $\ref{EW}$ and $\ref{F}$ (Blue symbols for H$\alpha$ + [N$\scriptstyle\rm II$] parameters and red symbols for H$\alpha$-only parameters).  }
\vspace{\baselineskip}
\begin{tabular}{lllrr} \hline
\\
\multicolumn{1}{l}{Ref. } &
\multicolumn{1}{l}{Papers } &
\multicolumn{1}{l}{Authors} &
\multicolumn{1}{r}{H$\alpha$ Types} &
\multicolumn{1}{r}{Symbols}\\
\\
\hline
\\

1 & 2002, A$\&$A, 384, 383 & Iglesias-Paramo  et al. & H$\alpha$ + [N$\scriptstyle\rm II$] & open triangle\\
2 & 1998, MNRAS, 300, 205 & Moss et al. & H$\alpha$ + [N$\scriptstyle\rm II$] & open circle \\
3 & 1984, AJ, 89, 1279 & Kennicutt  et al. & H$\alpha$ + [N$\scriptstyle\rm II$] &  filled triangle \\
4 & 1998, AJ, 115, 1745 & Gavazzi  et al. & H$\alpha$ + [N$\scriptstyle\rm II$] & cross \\
5 & 2002, A$\&$A, 386, 114  & Gavazzi  et al. & H$\alpha$ + [N$\scriptstyle\rm II$]  & open square  \\
6 & 2003, AJ, 597, 210 & Gavazzi  et al. & H$\alpha$ + [N$\scriptstyle\rm II$] & open square\\
7 & 2006, A$\&$A, 446, 839 & Gavazzi  et al. & H$\alpha$ + [N$\scriptstyle\rm II$]  & open square\\
8 & 2002, ApJ, 578, 842 & Sakai et al. & H$\alpha$  & filled square \\
9 & 2008, in preparation & Sakai et al. & H$\alpha$ & filled circle \\

\\
\hline
\\
\end{tabular}
\end{table*}

\subsection{Calculation of H$\alpha$ fluxes}

H$\alpha$  fluxes are determined using the following equations \citep{GAVA06}:

\begin{equation}
f_\alpha = CF \times  Z_f \times \left[\frac{C_{H\alpha}}{t \times T_{n}(6563(1+z))}\right]
\label{F1}
\end{equation}

and

\begin{equation}
CF =  1 + \frac{\int T_{n}(\lambda)d\lambda}{\int T_{b}(\lambda)d\lambda  }
\label{F2}
\end{equation}

where $Z_f$ is the calibrated zero point for H$\alpha$ emission line fluxes in units of erg~s$^{-1}$~cm$^{-2}$ and $t$ is the integration time in seconds for the narrow band H$\alpha$ images. $T_n$  and  $T_b$ are the transmissivities of narrow band and broad band filters, respectively.  $T_n(6563(1+z))$ was used in the same way in the calculation of equivalent widths. Because the broad band images used for continuum subtraction also included the emission line fluxes, potentially resulting in oversubtraction, an appropriate correction was applied to the $f_{\alpha}$ value. This gives a small change in the correcting factor, $CF$ in equation $\ref{F2}$.

Nine papers, listed in table $\ref{EWLit}$, have reported H$\alpha$ emission line fluxes of 18 galaxies in the sample of this study, whether including [N$\scriptstyle\rm II$] line fluxes or not. Only 13 galaxies have published total H$\alpha$ + [N$\scriptstyle\rm II$] emission line fluxes; these are in the first 7 of these papers. These galaxies were used to calibrate the flux zero-point. For galaxies which have published fluxes in more than one paper, the H$\alpha$ fluxes were averaged.

Firstly, the primary flux zero-point, $Z_f$ was set to 1.0 and instrumental H$\alpha$ fluxes, $f_{\alpha,Instr}$ were calculated using equation $\ref{F1}$. Calibrated flux zero-points, $Z_F$ were determined by dividing literature H$\alpha$ fluxes, $f_{\alpha,Lit}$ by the instrumental H$\alpha$ fluxes. The averaged zero-point was 3.32$\times$10$^{-16}$ erg s$^{-1}$ cm$^{-2}$ with standard error of 3.14$\times$10$^{-17}$ erg s$^{-1}$ cm$^{-2}$.

The parameters of this calibration are listed in table $\ref{FZ}$ as follows: Column 1, SDSS identifications with J2000 right ascension and declination; Columns 2 and 3 are logarithmic and decimal H$\alpha$~+~[N$\scriptstyle\rm II$] fluxes in units of erg s$^{-1}$ cm$^{-2}$, respectively; Column 4, the reference number of the paper as listed in table $\ref{EWLit}$; Columns 5 and 6 are the continuum-subtracted  H$\alpha$~+~[N$\scriptstyle\rm II$] flux counts in units of ADUs and their signal to noise ratios, respectively; Column 7, the instrumental H$\alpha$~+~[N$\scriptstyle\rm II$] fluxes; and Column 8 contains the calibrated flux zero-point of each galaxy, plus the overall median, mean, standard deviation and standard error.

\begin{table*}
\begin{center}
\caption{\label{FZ} Flux zero-point calibration }
\vspace{\baselineskip}
\begin{tabular}{lrrrrrrrr} \hline

\\
\multicolumn{1}{l}{Name } &
\multicolumn{1}{r}{$Log  f_{\alpha,Lit}$ } &
\multicolumn{1}{r}{$f_{\alpha,Lit}$} &
\multicolumn{1}{r}{Ref.} &
\multicolumn{1}{r}{$C_{H\alpha}$}& 
\multicolumn{1}{r}{$S/N$}& 
\multicolumn{1}{r}{$f_{\alpha,Instr}$}& 
\multicolumn{1}{r}{$Z_f$}& \\
\\
\hline
\\

J114239.31+195808.1 &-13.89& 1.29$\times$10$^{-14}$ &2&10001&53&40&3.24$\times$10$^{-16}$& \\ 
J114240.34+195717.0 &-14.7& 2.00$\times$10$^{-15}$ &1&2427&31&10&2.07$\times$10$^{-16}$& \\ 
J114256.45+195758.3 &-12.81& 1.55$\times$10$^{-13}$ & 1,2,3,4 &99819&82&397&3.9$\times$10$^{-16}$& \\ 
J114313.09+200017.3 &-12.65& 2.2$\times$x10$^{-13}$ & 1,2,3,4 &97929&105&394&5.69$\times$10$^{-16}$& \\ 
J114341.54+200137.0 &-14.15& 7.08$\times$10$^{-15}$ &1&4448&19&20&3.49$\times$10$^{-16}$& \\ 
J114348.89+201454.0 &-12.805& 1.57$\times$10$^{-13}$ & 1,5 &68247&152&389&4.03$\times$10$^{-16}$& \\ 
J114349.07+195806.4 &-12.19& 6.46$\times$10$^{-13}$ &1&604343&48&2519&2.56$\times$10$^{-16}$& \\ 
J114349.87+195834.8 &-13.99& 1.02$\times$10$^{-14}$ &1&10169&16&40&2.53$\times$10$^{-16}$& \\ 
J114358.23+201107.9 &-13.1& 7.94$\times$10$^{-14}$ &1&48729&33&226&3.51$\times$10$^{-16}$& \\ 
J114358.95+200437.2 &-12.8& 1.58$\times$10$^{-13}$ & 1,2,3 &163009&64&648&2.44$\times$10$^{-16}$& \\ 
J114447.03+200730.3 &-13.32& 4.79$\times$10$^{-14}$ &7&90798&13&394&1.21$\times$10$^{-16}$& \\ 
J114447.79+194624.3 &-13.11& 7.76$\times$10$^{-14}$ & 1,2,6 &57554&76&230&3.38$\times$10$^{-16}$& \\ 
J114451.09+194718.2 &-15.11& 7.76$\times$10$^{-16}$ &6&384&35&2&5.14$\times$10$^{-16}$& \\ 

&&&&&&&& \\ 
 Median &&&&&&&3.38$\times$10$^{-16}$& \\ 
 Mean &&&&&&&3.32$\times$10$^{-16}$& \\ 
 Std. dev. &&&&&&&1.22$\times$10$^{-16}$& \\ 
 Std. err. &&&&&&& 3.14$\times$10$^{-17}$ & \\

\\
\hline

\end{tabular}
\end{center}
\end{table*}

\begin{figure}
\centering
\vspace*{.6cm}
\includegraphics[width=0.35\textwidth,height=0.35\textheight,angle=-90]{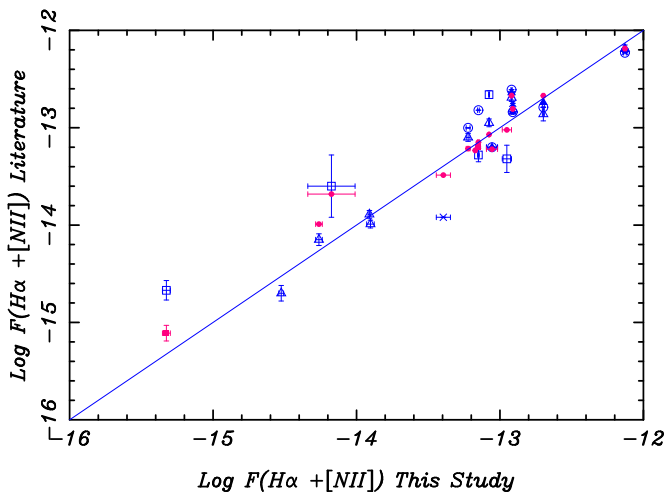}
\caption{Comparison of the H$\alpha$ emission line fluxes between the literature and this study. The symbols are listed in table $\ref{EWLit}$. }
\label{F}
\end{figure}

Figure $\ref{F}$ shows a comparison of the H$\alpha$ emission line fluxes between the literature and this study, along with the one-to-one line, with the same symbols as Fig. $\ref{EW}$.  Blue symbols represent H$\alpha$ + [N$\scriptstyle\rm II$] fluxes of the 13 calibrator galaxies and red symbols represent the galaxies which have published H$\alpha$ fluxes. Some galaxies have flux values from more than one paper, with or without flux errors.

Some scatter is seen in the comparisons of both H$\alpha$ equivalent widths and fluxes between the literature and this study (Figs. $\ref{EW}$ and $\ref{F}$). One possible source of scatter is differences in the apertures used between this study and the literature.  Substantial errors can also result from continuum subtraction, particularly for galaxies with low EW values.  Overall, the agreement between our values and those from the literature is good, with no evidence for systematic biases, and the scatter is similar to that found in other comparisons between studies. 

The result of H$\alpha$ + [N$\scriptstyle\rm II$] equivalent widths and fluxes of 72 cluster galaxies are listed in the H$\alpha$ emission catalogue, table $\ref{PhotCat}$. Column 1 is SDSS identifications. Columns 2 -- 4 show the morphological types, T-types and disturbance. Columns 5 -- 6 are recession velocities and references. Columns 7 -- 8 are H$\alpha$ + [N$\scriptstyle\rm II$] equivalent widths and fluxes. Table $\ref{avgf}$ shows 24 emission line galaxies in our sample, that were also reported in literature. Columns 2 -- 5 are available H$\alpha$ + [N$\scriptstyle\rm II$] fluxes from literature as follows: \citet{GAVA98,GAVA02,GAVA03B,GAVA06}, \citet{KENN84}, \citet{IGLE02} and \citet{THOM08}, respectively. Column 6 is the H$\alpha$ + [N$\scriptstyle\rm II$] fluxes of this study. Column 7 shows the mean fluxes and errors of H$\alpha$ + [N$\scriptstyle\rm II$] from the literature and this study.

\section{Summary}
\label{}

We present new photometric and spectroscopic data of the galaxy population of cluster Abell 1367, covering the cluster centre up to about 1.0 Abell  radius ($r_A$) or $\sim$ 2.2 Mpc to the north direction. The data are useful to investigate the photometric properties and star formation activity of galaxies in Abell 1367. This paper summarises the photometric observations and data reduction for $U$, $B$, $R$ and H$\alpha$ bands using WFC on the INT, and $J$ and $K$ bands using WFCAM on UKIRT. Spectroscopic data from the WYFFOS multifibre spectrograph on the WHT are also presented, along with derived recession velocities. A galaxy sample was selected using a limiting $B$-band apparent magnitude and minimum FWHM to limit contaminated background and foreground galaxies. The selected sample contains 303 galaxies, for which we present a photometric catalogue (table $\ref{Photom}$). Our new spectroscopic data for 84 galaxies are presented in a spectroscopic catalogue (table $\ref{SpecCat}$); after combining with published spectra, a total of 126 of our galaxies have measured recession velocities. Of these, 72 galaxies, with spectroscopically confirmed cluster membership, have measures of star formation activity in terms of H$\alpha$ equivalent widths and fluxes, which are presented in a H$\alpha$ emission catalogue (table $\ref{PhotCat}$).

\begin{acknowledgements}

We thank the Cambridge Astronomical Survey Unit (CASU), Institute of Astronomy, Cambridge University, for the pipeline processing of the INT and UKIRT images. This research has made use of the NASA/IPAC Extragalactic Database (NED), operated by the Jet Propulsion Laboratory, California Institute of Technology, under contract with the National Aeronautics and Space Administration.  The referee is thanked for many helpful comments regarding the content and presentation of this paper.  WK, PAJ and DC dedicate this paper to the memory of their respected and sadly-missed colleague Chris Moss, who died on the 12th May 2010.

\end{acknowledgements}



\begin{table*}
\tiny
\caption{\label{SpecCat}Redshift measurements for galaxies in the region of Abell 1367}
\begin{tabular}{@{}lrl@{}}
\hline
\hline

\multicolumn{1}{l}{SDSS } &
\multicolumn{1}{c}{$cz$ } &
\multicolumn{1}{l}{Spectral Lines}  \\

\multicolumn{1}{l}{} &
\multicolumn{1}{c}{(km \rm s$^{-1}$)} &
\multicolumn{1}{l}{}  \\

\hline

 J114229.18+200713.7			   &   	6250~~~				   &  	H$\beta$~$\lambda$~4861,	[O$\scriptstyle\rm	III$]~$\lambda$~4959, 	[O$\scriptstyle\rm	III$]~$\lambda$~5007,	H$\alpha$~$\lambda$~6563,	[N$\scriptstyle\rm	II$]~$\lambda$~6583,	[S$\scriptstyle\rm	II$]~$\lambda$~6717,	[S$\scriptstyle\rm	II$]~$\lambda$~6731	   	\\ 
 J114231.83+203419.1			   &   	56941~~~				   &   	G~band~$\lambda$~4304,	MgI~$\lambda$~5175	   	\\				 
 J114232.60+200100.7			   &   	21098~~~			   &   		[O$\scriptstyle\rm	III$]~$\lambda$~4959,	[O$\scriptstyle\rm	III$]~$\lambda$~5007,	H$\alpha$~$\lambda$~6563, [S$\scriptstyle\rm	II$]~$\lambda$~6717,	[S$\scriptstyle\rm	II$]~$\lambda$~6731	   	\\	 
 J114234.59+200111.8			   &   	144302~~~				   &   		[O$\scriptstyle\rm	II$]~$\lambda$~3727, 		H$\beta$~$\lambda$~4861,	[O$\scriptstyle\rm	III$]~$\lambda$~5007   	\\		 
 J114237.58+200715.1			   &   	39925~~~				   &   	H$\alpha$~$\lambda$~6563,	[N$\scriptstyle\rm	II$]~$\lambda$~6583	   	\\					 
 J114238.29+194718.1			   &   	27303~~~				   &   	[N$\scriptstyle\rm	II$]~$\lambda$~6548,  H$\alpha$~$\lambda$~6563,	[N$\scriptstyle\rm	II$]~$\lambda$~6583	   	\\
\end{tabular}

\end{table*}

\setcounter{table}{7}
\begin{table*}
\tiny
\caption{{Continued.}}
\begin{tabular}{@{}lrl@{}}
\hline
\hline
\multicolumn{1}{l}{SDSS } &

\multicolumn{1}{c}{$cz$ } &
\multicolumn{1}{l}{Spectral Lines} \\

\multicolumn{1}{l}{} &

\multicolumn{1}{c}{(km \rm s$^{-1}$)} &
\multicolumn{1}{l}{}  \\

\hline

 J114448.04+193933.7			   &   	20402~~~				   &	[O$\scriptstyle\rm	III$]~$\lambda$~5007,   	H$\alpha$~$\lambda$~6563	   	\\							 
 J114448.90+194828.8			   &   	20343~~~				   &   	H$\alpha$~$\lambda$~6563,	[N$\scriptstyle\rm	II$]~$\lambda$~6583	   	\\							 
 J114449.16+194742.2			   &   	5579~~~				   &   		MgI~$\lambda$~5175,	NaD~$\lambda$~5893, H$\alpha$~$\lambda$~6563,	[N$\scriptstyle\rm	II$]~$\lambda$~6583	   	\\	
 J114450.19+204134.9			   &   	55694~~~				   &   	G~band~$\lambda$~4304,	MgI~$\lambda$~5175,	FeI~$\lambda$~5269,	NaD~$\lambda$~5893	   	\\ 
 J114450.29+203023.9			   &   	83108~~~				   &   	[O$\scriptstyle\rm	II$]~$\lambda$~3727, H$\beta$~$\lambda$~4861,	[O$\scriptstyle\rm	III$]~$\lambda$~4959,	[O$\scriptstyle\rm	III$]~$\lambda$~5007	   	\\	 
 J114450.31+195903.7			   &   	4900~~~				   &   	H$\alpha$~$\lambda$~6563,	[N$\scriptstyle\rm	II$]~$\lambda$~6583	   	\\									 
 J114451.19+202333.9			   &   	26625~~~				   &   	H$\beta$~$\lambda$~4861,	[O$\scriptstyle\rm	III$]~$\lambda$~4959,	[O$\scriptstyle\rm	III$]~$\lambda$~5007, 	[N$\scriptstyle\rm	II$]~$\lambda$~6548,  H$\alpha$~$\lambda$~6563,		[N$\scriptstyle\rm	II$]~$\lambda$~6583	   	\\ 
 J114451.81+210001.6			   &   	53973~~~				   &   	H$\gamma$~$\lambda$~4340,  H$\beta$~$\lambda$~4861,	[O$\scriptstyle\rm	III$]~$\lambda$~4959,		[O$\scriptstyle\rm	III$]~$\lambda$~5007	   	\\

\hline
\end{tabular}
\\
\begin{tabular}{l}
\\
{\it Notes to Table~\ref{SpecCat}:}
Column 1: SDSS identifications.
Column 2: recession velocities.
Column 3: spectral lines.

\end{tabular}

\end{table*}

\begin{table*}
\tiny
\caption{Photometric measurements for galaxies in the surveyed region of Abell 1367}
\label{Photom}

\begin{tabular}{lrrccccc}
\hline
\hline
\multicolumn{1}{l}{SDSS } &
\multicolumn{1}{c}{$D_{24}$ } &
\multicolumn{1}{c}{$d_{24}$ } &
\multicolumn{1}{c}{$m_U$} &
\multicolumn{1}{c}{$m_B$} &
\multicolumn{1}{c}{$m_R$} &
\multicolumn{1}{c}{$m_J$} &
\multicolumn{1}{c}{$m_K$}  \\

\multicolumn{1}{l}{ } &
\multicolumn{1}{c}{($^{\prime}$$^{\prime}$) } &
\multicolumn{1}{c}{($^{\prime}$$^{\prime}$) } &
\multicolumn{1}{c}{} &
\multicolumn{1}{c}{} &
\multicolumn{1}{c}{} &
\multicolumn{1}{c}{} &
\multicolumn{1}{c}{}  \\

\hline

 J114228.63+194451.4		&		5.8~		&		3.0~		&	20.77	(	0.05	)	&	20.74	(	0.06	)	&	18.96	(	0.05	)	&	16.95	(	0.02	)	&	15.56	(	0.02	)	\\
 J114229.07+195237.3		&		9.4~		&		8.4~		&	18.90	(	0.03	)	&	19.07	(	0.03	)	&	17.39	(	0.03	)	&	15.95	(	0.02	)	&	14.72	(	0.02	)	\\
 J114229.18+200713.7		&		13.8~		&		5.8~		&	19.46	(	0.04	)	&	19.62	(	0.05	)	&	18.76	(	0.07	)	&	17.79	(	0.10	)	&	17.33	(	0.12	)	\\
 J114230.48+200738.5		&		2.8~		&		1.2~		&	20.96	(	0.08	)	&	21.20	(	0.10	)	&	20.36	(	0.17	)	&	19.98	(	0.45	)	&	18.99	(	0.35	)	\\
 J114232.50+200102.3		&		9.2~		&		4.2~		&	19.29	(	0.04	)	&	19.62	(	0.05	)	&	18.67	(	0.05	)	&	17.77	(	0.04	)	&	16.92	(	0.04	)	\\
 J114233.47+205523.8		&		19.6~		&		7.2~		&	20.04	(	0.08	)	&	19.89	(	0.06	)	&	18.56	(	0.07	)	&					&					\\
\end{tabular}

\end{table*}

\setcounter{table}{8}
\begin{table*}
\tiny
\caption{{Continued.}}
\begin{tabular}{lrrccccc}
\hline
\hline
\multicolumn{1}{l}{SDSS } &
\multicolumn{1}{c}{$D_{24}$ } &
\multicolumn{1}{c}{$d_{24}$ } &
\multicolumn{1}{c}{$m_U$} &
\multicolumn{1}{c}{$m_B$} &
\multicolumn{1}{c}{$m_R$} &
\multicolumn{1}{c}{$m_J$} &
\multicolumn{1}{c}{$m_K$}  \\

\multicolumn{1}{l}{ } &
\multicolumn{1}{c}{($^{\prime}$$^{\prime}$) } &
\multicolumn{1}{c}{($^{\prime}$$^{\prime}$) } &
\multicolumn{1}{c}{} &
\multicolumn{1}{c}{} &
\multicolumn{1}{c}{} &
\multicolumn{1}{c}{} &
\multicolumn{1}{c}{}  \\

\hline

 J114328.57+195336.4 		&		12.4~		&		7.8~		&	17.84	(	0.03	)	&	17.96	(	0.02	)	&	16.60	(	0.03	)	&	15.10	(	0.02	)	&	13.99	(	0.02	)	\\
 J114329.65+202159.7 		&		11.8~		&		8.2~		&	18.94	(	0.04	)	&	19.17	(	0.04	)	&	17.58	(	0.04	)	&	15.94	(	0.02	)	&	14.67	(	0.02	)	\\
 J114330.30+205541.5 		&		8.0~		&		4.8~		&	19.67	(	0.05	)	&	19.51	(	0.04	)	&	17.77	(	0.03	)	&					&					\\
 J114330.36+205301.9		&		12.2~		&		3.8~		&	21.32	(	0.23	)	&	21.25	(	0.19	)	&	19.42	(	0.17	)	&					&					\\
 J114330.65+201719.7 		&		1.6~		&		0.8~		&	21.61	(	0.09	)	&	21.51	(	0.10	)	&	20.47	(	0.14	)	&	19.30	(	0.10	)	&	18.73	(	0.14	)	\\
 J114331.08+200517.8 		&		6.0~		&		5.0~		&	20.55	(	0.06	)	&	20.50	(	0.06	)	&	18.98	(	0.07	)	&	17.77	(	0.05	)	&	16.83	(	0.05	)	\\
\end{tabular}

\end{table*}

\setcounter{table}{8}
\begin{table*}
\tiny
\caption{{Continued.}}
\begin{tabular}{lrrccccc}
\hline
\hline
\multicolumn{1}{l}{SDSS } &
\multicolumn{1}{c}{$D_{24}$ } &
\multicolumn{1}{c}{$d_{24}$ } &
\multicolumn{1}{c}{$m_U$} &
\multicolumn{1}{c}{$m_B$} &
\multicolumn{1}{c}{$m_R$} &
\multicolumn{1}{c}{$m_J$} &
\multicolumn{1}{c}{$m_K$}  \\

\multicolumn{1}{l}{ } &
\multicolumn{1}{c}{($^{\prime}$$^{\prime}$) } &
\multicolumn{1}{c}{($^{\prime}$$^{\prime}$) } &
\multicolumn{1}{c}{} &
\multicolumn{1}{c}{} &
\multicolumn{1}{c}{} &
\multicolumn{1}{c}{} &
\multicolumn{1}{c}{}  \\

\hline

 J114356.65+194014.2		&		7.2~		&		5.2~		&	19.01	(	0.04	)	&	18.84	(	0.04	)	&	17.48	(	0.04	)	&	16.27	(	0.02	)	&	15.32	(	0.02	)	\\
 J114356.67+200150.9 		&		5.6~		&		5.4~		&	18.99	(	0.03	)	&	18.69	(	0.03	)	&	17.09	(	0.03	)	&	15.60	(	0.02	)	&	14.74	(	0.02	)	\\
 J114356.84+195720.8 		&		3.4~		&		3.2~		&	18.14	(	0.03	)	&	17.73	(	0.03	)	&	16.08	(	0.03	)	&	14.49	(	0.02	)	&	13.56	(	0.02	)	\\
 J114356.97+195000.0 		&		6.6~		&		5.8~		&	20.00	(	0.05	)	&	19.76	(	0.05	)	&	18.14	(	0.06	)	&	16.78	(	0.03	)	&	16.01	(	0.02	)	\\
 J114357.01+202238.3 		&		20.2~		&		9.6~		&	17.53	(	0.03	)	&	17.17	(	0.03	)	&	15.46	(	0.03	)	&	13.62	(	0.02	)	&	12.54	(	0.02	)	\\
 J114357.44+195608.4 		&		12.4~		&		7.8~		&	17.86	(	0.03	)	&	17.53	(	0.03	)	&	16.01	(	0.02	)	&	14.56	(	0.02	)	&	13.64	(	0.02	)	\\
\end{tabular}

\end{table*}

\setcounter{table}{8}
\begin{table*}
\tiny
\caption{{Continued.}}
\begin{tabular}{lrrccccc}
\hline
\hline

\multicolumn{1}{l}{SDSS } &
\multicolumn{1}{c}{$D_{24}$ } &
\multicolumn{1}{c}{$d_{24}$ } &
\multicolumn{1}{c}{$m_U$} &
\multicolumn{1}{c}{$m_B$} &
\multicolumn{1}{c}{$m_R$} &
\multicolumn{1}{c}{$m_J$} &
\multicolumn{1}{c}{$m_K$}  \\

\multicolumn{1}{l}{ } &
\multicolumn{1}{c}{($^{\prime}$$^{\prime}$) } &
\multicolumn{1}{c}{($^{\prime}$$^{\prime}$) } &
\multicolumn{1}{c}{} &
\multicolumn{1}{c}{} &
\multicolumn{1}{c}{} &
\multicolumn{1}{c}{} &
\multicolumn{1}{c}{}  \\

\hline

 J114422.63+202439.7 		&		21.6~		&		6.8~		&	19.20	(	0.04	)	&	19.21	(	0.04	)	&	17.45	(	0.04	)	&	15.68	(	0.02	)	&	14.30	(	0.02	)	\\
 J114422.80+202433.7 		&		13.2~		&		8.8~		&	19.15	(	0.04	)	&	18.93	(	0.04	)	&	17.54	(	0.04	)	&	16.35	(	0.02	)	&	15.59	(	0.03	)	\\
 J114422.92+202651.8 		&		8.2~		&		2.8~		&	20.30	(	0.06	)	&	20.57	(	0.08	)	&	19.08	(	0.06	)	&	17.53	(	0.08	)	&	16.34	(	0.07	)	\\
 J114423.19+202114.5 		&		6.2~		&		5.0~		&	20.01	(	0.05	)	&	19.79	(	0.05	)	&	17.60	(	0.03	)	&	15.79	(	0.02	)	&	14.61	(	0.02	)	\\
 J114423.78+195005.2 		&		6.0~		&		5.0~		&	20.54	(	0.05	)	&	20.06	(	0.05	)	&	18.50	(	0.04	)	&	17.12	(	0.03	)	&	16.18	(	0.02	)	\\
 J114425.10+200628.1		&		27.0~		&		24.2~		&	16.35	(	0.03	)	&	16.03	(	0.03	)	&	14.43	(	0.03	)	&	12.96	(	0.02	)	&	12.08	(	0.02	)	\\
\hline
\end{tabular}
\\
\begin{tabular}{l}

\\

{\it Notes to Table~\ref{Photom}:}
Column 1: SDSS identifications.
Columns 2 -- 3: Major and minor diameters at approximate $R_{24}$ isophotes. \\
Columns 4 -- 8: $UBRJK$ magnitudes and errors.

\end{tabular}

\end{table*}

\begin{table*}
\tiny
\caption{H$\alpha$ emission catalogue of galaxies in Abell 1367}
\label{PhotCat}

\begin{tabular}{llrcrcrr}
\hline
\hline

\multicolumn{1}{l}{SDSS } &
\multicolumn{1}{l}{Type } &
\multicolumn{1}{r}{T-Type} &
\multicolumn{1}{c}{Disturbance} &
\multicolumn{1}{c}{$cz$ } &
\multicolumn{1}{c}{Ref.} &
\multicolumn{1}{c}{$EW$(H$\alpha$ + [N$\scriptstyle\rm II$])} &
\multicolumn{1}{c}{Log $f$(H$\alpha$ + [N$\scriptstyle\rm II$])} \\

\multicolumn{1}{l}{} &
\multicolumn{1}{l}{ } &
\multicolumn{1}{r}{} &
\multicolumn{1}{c}{} &
\multicolumn{1}{c}{(km \rm s$^{-1}$) } &
\multicolumn{1}{c}{} &
\multicolumn{1}{c}{(\rm \AA)} &
\multicolumn{1}{c}{(erg \rm s$^{-1}$ cm$^{-2}$)} \\
\hline

  J114229.18+200713.7	&	                 Irr                 	&	10~~~~	&	                         -                         	&	6250~~~	&	                 *                 	&	62~~~~~~~~~	&	--14.60~~~~~~~~~	\\
  J114239.31+195808.1	&	                 Sc                 	&	5~~~~	&	                 Disturbed                 	&	7426~~~	&	                 *                 	&	56~~~~~~~~~	&	--13.91~~~~~~~~~	\\
  J114240.34+195717.0 	&	                 S0                 	&	--2~~~~	&	                         -                         	&	7586~~~	&	                 *                 	&	33~~~~~~~~~	&	--14.52~~~~~~~~~	\\
  J114246.11+195654.6 	&	                 E                 	&	--5~~~~	&	                         -                         	&	6216~~~	&	1	&		&		\\
  J114251.93+195657.0 	&	                 S0/a                 	&	0~~~~	&	                         -                         	&	5936~~~	&	2	&		&		\\
  J114253.19+201045.9 	&	                 S0/a                 	&	0~~~~	&	                         -                         	&	6261~~~	&	3	&		&		\\
\hline
\end{tabular}
\\
\begin{tabular}{l}
\\

{\it Notes to Table~\ref{PhotCat}:} 
Column 1: SDSS identifications. 
Columns 2 -- 4: Hubble types, T-types and disturbances, respectively.
Column 5: \\recession velocities.
Column 6: reference numbers of published papers, same as table $\ref{complst}$ and asterisk symbol for this study.\\ 
Columns 7 -- 8: equivalent widths and fluxes of H$\alpha$ + [N$\scriptstyle\rm II$], respectively.

\end{tabular}
\end{table*}

\begin{table*}
\begin{center}
\caption{\label{avgf} Mean fluxes and errors of H$\alpha$ + [N$\scriptstyle\rm II$] from the literature and this study.    }
\begin{tabular}[]{lcccccc} 
\hline
\hline

SDSS &&&  Log f(H$\alpha$ + [N$\scriptstyle\rm II$]) &&&  Log $<$f(H$\alpha$ + [N$\scriptstyle\rm II$])$>$ \\
&&&  (erg \rm s$^{-1}$ cm$^{-2}$) &&& (erg \rm s$^{-1}$ cm$^{-2}$)\\

\cline{2-6} \\
& (1) & (2) & (3) & (4) &  This Study &  \\
\hline

J114229.18+200713.7	&	   	&	   	&	   	&	   	&	--14.60    	&	--14.60    	(0.10)    	\\
J114239.31+195808.1	&	   	&	   	&	--13.89    	&	   	&	--13.91    	&	--13.90    	(0.07)    	\\
J114240.34+195717.0 	&	   	&	   	&	--14.70    	&	   	&	--14.52    	&	--14.61    	(0.07)    	\\
J114256.45+195758.3 	&	--12.75    	&	--12.84    	&	--12.81    	&	--12.74    	&	--12.91    	&	--12.81    	(0.03)    	\\
\hline
\end{tabular}
\end{center}

\begin{tabular}{l}

\\

\it Notes to Table $\ref{avgf}$: 
\rm Column 1: SDSS identifications. Column 2-5: fluxes of H$\alpha$ + [N$\scriptstyle\rm II$] from literature as follow: Gavazzi \\et al. (1998, 2002, 2003b, 2006), Kennicutt et al. (1984), Iglesias-Paramo et al. (2002) and Thomas et al. (2008), respectively. \\
Column 6: fluxes of H$\alpha$ + [N$\scriptstyle\rm II$] from this study. Column 7: mean fluxes and errors of H$\alpha$ + [N$\scriptstyle\rm II$].

\end{tabular}

\end{table*}


\begin{thebibliography}{}

\bibitem[{{Abell}(1958)}]{ABEL58}
    {Abell}, G.~O. 1958, ApJS, 3, 112

\bibitem[{{Bertin} \& {Arnouts}(1996)}]{BERT96}
    {Bertin}, E., \& {Arnouts}, S. 1996, A\&AS, 117, 393

\bibitem[{{Boselli}(1994)}]{BOSE94} 
    Boselli, A.\ 1994, A\&A, 292, 1 

\bibitem[{Boselli} {et al.}(2006)]{BOSEEA06} 
    Boselli, A., Boissier, S., Cortese, L., Gil de Paz, A., Seibert, M., 
    Madore, B.~F., Buat, V., \& Martin, D.~C. 2006, ApJ, 651, 811


\bibitem[{{Boselli} \& {Gavazzi}(2006)}]{BOSE06}
    {Boselli}, A., \& {Gavazzi}, G. 2006, PASP, 118, 517

\bibitem[{{Boselli} {et al.}(2002)}]{BOSE02} 
    Boselli, A., Lequeux, J., \& Gavazzi, G. 2002, A\&A, 384, 33 

\bibitem[{{Buta}(1996)}]{BUTA96} {Buta}, R. 1996, AJ, 111, 591

\bibitem[{{Byrd} \& {Valtonen}(1990)}]{BYRD90}
    {Byrd}, G., \& {Valtonen}, M. 1990, ApJ, 350, 89

\bibitem[{{Cardelli} {et~al.}(1989)}]{CARD89}
    {Cardelli}, J.~A., {Clayton}, G.~C., \& {Mathis}, J.~S. 1989,
    ApJ, 345, 245

\bibitem[{{Condon} {et~al.}(1982)}]{COND82}
    {Condon}, J.~J., {Condon}, M.~A., {Gisler}, G., \& {Puschell}, J.~J. 1982,
    ApJ, 252, 102

\bibitem[{{Cortese} {et~al.}(2004)}]{CORT04}
    Cortese, L., Gavazzi, G., Boselli, A., Iglesias-Paramo, J., \& Carrasco, L. 2004,
    A\&A, 425, 429

\bibitem[{{Cortese} {et~al.}(2003)}]{CORT03}
    Cortese, L., Gavazzi, G., Iglesias-Paramo, J., Boselli, A., \& Carrasco, L. 2003,
    A\&A, 401, 471

\bibitem[{{Cayatte} {et al.}(1994)}]{CAYA94} 
    Cayatte, V., Kotanyi, C., Balkowski, C., \& van Gorkom, J.~H. 1994, AJ, 107, 1003 

\bibitem[{{de Vaucouleurs \& Longo}(1988)}]{DEVA88} de Vaucouleurs, G., \& Longo, G. 1988,
    Catalogue of visual and infrared photometry of galaxies (1961-1985)

\bibitem[{{Donnelly} {et al.}(1998)}]{DONN98}
    Donnelly, R. H., Markevitch, M., Forman, W. et al. 1998, ApJ, 500, 138

\bibitem[{Dressler}(2004)]{DRES04} Dressler, A. 2004, Clusters 
of Galaxies: Probes of Cosmological Structure and Galaxy Evolution, 206 

\bibitem[{{Fujita}(1998)}]{FUJI98} {Fujita}, Y. 1998, ApJ, 509, 587

\bibitem[{{Finn} {et al.}(2005)}]{FINN05}
    Finn, R. A., Zaritsky, D., McCarthy, D. W., Jr., Poggianti, B., Rudnick, G., Halliday, C., Milvang-Jensen, B., Pello, R., \& Simard, L. 2005, 
    ApJ, 630, 206


\bibitem[{{Gavazzi}(1989)}]{GAVA89} 
    Gavazzi, G. 1989, ApJ, 346, 59 

\bibitem[{{Gavazzi} {et al.}(2006)}]{GAVA06}
    Gavazzi, G., Boselli, A., Cortese, L., Arosio, I., Gallazzi, A., Pedotti, P., \& Carrasco, L. 2006,
    A\&A, 446, 839

\bibitem[{{Gavazzi} {et al.}(2003a)}]{GAVA03A}
    Gavazzi, G., Boselli, A., Donati, A., Franzetti, P., \& Scodeggio, M. 2003a,
    A\&A, 400, 451

\bibitem[{{Gavazzi} {et al.}(2002)}]{GAVA02}
    Gavazzi, G., Boselli, A., Pedotti, P., Gallazzi, A., \& Carrasco, L. 2002,
    A\&A, 386, 114

\bibitem[{{Gavazzi} {et al.}(1998)}]{GAVA98}
    Gavazzi, G., Catinella, B., Carrasco, L. , Boselli, A., \& Contursi, A. 1998, AJ, 115, 1745

\bibitem[{{Gavazzi} {et al.}(2003b)}]{GAVA03B}
    Gavazzi, G., Cortese, L., Boselli, A., Iglesias-Paramo, J., Vilchez, J., \& Carrasco, L. 2003b,
    AJ, 597, 210

\bibitem[{{Gunn} \& {Gott}(1972)}]{GUNN72}
{Gunn}, J.~E., \& {Gott}, J.~R.~I. 1972, ApJ, 176, 1

\bibitem[{{Haynes} \& {Giovanelli}(1984)}]{HAYN84} 
    Haynes, M.~P., \& Giovanelli, R. 1984, AJ, 89, 758 

\bibitem[{{Haynes} {et al.}(1997)}]{HAYN97}
    Haynes, M. P., Giovanelli, R., Herter, T., Vogt, N. P.,
    Freudling, W., Maia, M. A. G., Salzer, J. J., \& Wegner, G. 1997, AJ, 113, 1197

\bibitem[{{Henriksen} \& {Byrd}(1996)}]{HENR96}
{Henriksen}, M., \& {Byrd}, G. 1996, ApJ, 459, 82

\bibitem[{{Iglesias-Paramo} {et al.}(2002)}]{IGLE02}
    Iglesias-Paramo, J., Boselli, A., Cortese, L., Vilchez, J. M., \& Gavazzi, G. 2002, A\&A, 384, 383

\bibitem[{{Iono} {et~al.}(2004)}]{IONO04}
{Iono}, D., {Yun}, M.~S., \& {Mihos}, J.~C. 2004, ApJ, 616, 199

\bibitem[{{Jarrett et al.}(2003)}]{JARR03} Jarrett T.~H., Chester T., Cutri R.,
Schneider S.~E., Huchra J.~P., 2003, AJ, 125, 525 

\bibitem[{{Jarrett et al.}(2000)}]{JARR00} Jarrett T.~H., Chester T., Cutri R.,
Schneider S., Skrutskie M., Huchra J.~P., 2000, AJ, 119, 2498 

\bibitem[{{Keel} {et~al.}(1985)}]{KEEL85}
{Keel}, W.~C., {Kennicutt}, R.~C., {Hummel}, E., \& {van der Hulst}, J.~M.
  1985, AJ, 90, 708

\bibitem[{{Kennicutt}(1983a)}]{KENN83A} Kennicutt, R. C. 1983a, ApJ, 272, 54

\bibitem[{{Kennicutt}(1983b)}]{KENN83B} Kennicutt, R.~C. 1983b, AJ, 88, 483 

\bibitem[{{Kennicutt}(1998)}]{KENN98} Kennicutt, R. C. 1998, ARA\&A, 36, 189

\bibitem[{{Kennicutt} {et al.}(1984)}]{KENN84}
Kennicutt, R. C., Bothun, G. D., \& Schommer, R. A. 1984, AJ, 89, 1279

\bibitem[{{Kennicutt} {et~al.}(1987)}]{KENN87}
{Kennicutt}, R.~C., {Roettiger}, K.~A., {Keel}, W.~C., {van der Hulst}, J.~M.,
  \& {Hummel}, E. 1987, AJ, 93, 1011

\bibitem[{{Kennicutt} {et al.}(1994)}]{KENN94}
Kennicutt, R. C., Tamblyn, P., \& Congdon, C. E. 1994, ApJ, 435, 22

\bibitem[{{Koopmann} {et al.}(2006)}]{KOOP06} 
    Koopmann, R.~A., Haynes, M.~P., \& Catinella, B. 2006, AJ, 131, 716 

\bibitem[{{Koopmann} \& {Kenney}(1998)}]{KOOP98} 
    Koopmann, R.~A., \& Kenney, J.~D.~P. 1998, ApJL, 497, L75 

\bibitem[{{Koopmann} \& {Kenney}(2004a)}]{KOOP04A} 
    Koopmann, R.~A., \& Kenney, J.~D.~P. 2004a, ApJ, 613, 851 

\bibitem[{{Koopmann} \& {Kenney}(2004b)}]{KOOP04B} 
    Koopmann, R.~A., \& Kenney, J.~D.~P. 2004b, ApJ, 613, 866 

\bibitem[{{Mihos} {et~al.}(1992)}]{MIHO92}
{Mihos}, J.~C., {Richstone}, D.~O., \& {Bothun}, G.~D. 1992, ApJ, 400, 153

\bibitem[{{Moore} {et~al.}(1996)}]{MOOR96}
{Moore}, B., {Katz}, N., {Lake}, G., {Dressler}, A., \& {Oemler}, A. 1996, Nature, 379, 613

\bibitem[{{Moore} {et~al.}(1999)}]{MOOR99}
{Moore}, B., {Lake}, G., {Quinn}, T., \& {Stadel}, J. 1999, MNRAS, 304, 465

\bibitem[{{Moss} {et~al.}(1998)}]{MOSS98} Moss, C., Whittle, M., \& Pesce, J. E. 1998,
    MNRAS, 300, 205

\bibitem[{{O'Donnell}(1994)}]{ODON94} O'Donnell, J. E. 1994, ApJ, 422, 1580

\bibitem[{{Poggianti}(1997)}]{POGG97} Poggianti, B. M. 1997, A\&AS, 122, 399

\bibitem[{{Rines} {et~al.}(2003)}]{RINE03} 
    Rines, K., Geller, M. J., Kurtz, M. J., \& Diaferio, A. 2003, AJ, 126, 2152

\bibitem[{{Robinson}(1965)}]{ROBI65} 
    Robinson, B.~J. 1965, Nature, 208, 993 

\bibitem[{{Sakai} {et~al.}(2002)}]{SAKA02} Sakai, S., Kennicutt, R. C., van der Hulst, J. M.,  \& Moss, C. 2002,
    ApJ, 578, 842

\bibitem[{{Schlegel} {et~al.}(1998)}]{SCHL98} Schlegel, D. J., Finkbeiner, D. P., \& Davis, M. 1998,
    ApJ, 500, 525

\bibitem[{{Smith} {et~al.}(2004)}]{SMIT04} 
    Smith, R. J., Hudson, M. J., Nelan, J. E., Moore, S. A. W., Quinney, S. J., Wegner, G. A., Lucey, J. R.,
    Davies, R. L., Malecki, J. J. Schade, D., \& Suntzeff, N. B. 2004, AJ, 128, 1558

\bibitem[{{Struble \& Rood}(1999)}]{STRU99} Struble, M. F., \& Rood, H. J. 1999,
    ApJS, 125, 35

\bibitem[{{Sullivan} {et~al.}(2000)}]{SULL00}
    Sullivan, M., Treyer, M. A., Ellis, R. S., Bridges, T. J., Milliard, B., \& Donas, J. 2000,
    MNRAS, 312, 442

\bibitem[{{Taylor} {et~al.}(2005)}]{TAYL05}
    Taylor, V. A., Jansen, R. A., Windhorst, R. A., Odewahn, S. C., \& Hibbard, J. E. 2005,
    ApJ, 630, 784

\bibitem[{{Thomas} {et~al.}(2008)}]{THOM08}
    Thomas, C. F., Moss, C., James, P. A., Bennett, S. M., Aragon-Salamanca, A., \& Whittle, M. 2008, A\&A, 486, 755

\bibitem[{{Warmels}(1988)}]{WARM88} 
    Warmels, R.~H. 1988, A\&AS, 72, 427

\bibitem[{{Zabludoff} {et~al.}(1990)}]{ZABL90}
    Zabludoff, A. I., Huchra J. P., \& Geller M. J. 1990, ApJS, 74, 1

\end{thebibliography}
\end{document}